\documentclass[12pt,preprint]{aastex}

\begin{document}

\title {AN OBSERVATIONAL SIGNATURE OF EVOLVED OCEANS ON EXTRA-SOLAR TERRESTRIAL PLANETS}

\author{M. Jura} 
\affil{Department of Physics and Astronomy, University of California,
    Los Angeles CA 90095-1562; jura@clotho.astro.ucla.edu}

\begin{abstract}
 
The increase in luminosity with time of a main sequence star eventually can lead to substantial
evaporation of the oceans  on an orbiting terrestrial planet.  Subsequently, the gas-phase H$_{2}$O in the planet's upper atmosphere can be photodissociated by stellar ultraviolet and the resulting atomic hydrogen  then may be lost  in a wind. This gaseous envelope may pass in front of the host star and 
  produce transient, detectable ultraviolet absorption in the Lyman lines in systems older than 1 Gyr.

\end{abstract}

\section{INTRODUCTION}
The possible existence  of life on other planets is of central interest in  modern astronomy.  A standard working hypothesis is that liquid
water is required for life, and here we describe an observational signature of evolved oceans on extra-solar terrestrial planets.   

The occultation of HD 209458 by a Jovian planet 
has demonstrated that it is possible to study the atmospheres of extra-solar planets with absorption line spectroscopy during a transit (Charbonneau et al. 2000, 2002).  The Lyman ${\alpha}$  spectrum of HD 209458 observed during its companion planet's transit can be explained 
 as an outflow of hydrogen from this planet (Vidal-Madjar et al. 2003. Liang et al. 2003) that also has entrained other  gases (Vidal-Madjar et al. 2004).
Below, we describe an extension of this technique to study gaseous winds from terrestrial planets.    

To date, only Jovian mass extra-solar planets have been detected. The discovery of terrestrial planets by occultations of their host stars is
the goal of future space missions such as Kepler (see, for example, Jenkins 2002).  In addition to transits discovered by photometry, it may be possible to perform  transient absorption line
spectroscopy to study  gaseous outflows from terrestrial planets, and we suggest that atomic
hydrogen originating from evolved oceans might be detectable by this method.   
It seems likely that Venus once had oceans that were lost through a wind  (Watson, Donahue \& Walker 1981, Kasting \& Pollock 1983), and computing the structure and composition of a gaseous outflow  from an analog to Venus in orbit around a  star with an age less than 1 Gyr is a goal of models being developed  by Parkinson et al. (2003). 
However, as described below, the detection of an outflow from an Earth-like planet orbiting a star older than 1 Gyr is more promising.

Although not necessarily true in the past (see Sackmann \& Boothroyd 2003), currently, the Sun's luminosity is increasing with time  (see Girardi et al. 2000). Eventually, if significantly vaporized,  H$_{2}$O can 
dominate the composition of the Earth's atmosphere since the total mass of
the oceans is 1.4 ${\times}$ 10$^{24}$ g while the total mass of the current atmosphere is
5.1 ${\times}$ 10$^{21}$ g (Schubert \& Walterscheid 2000). In approximately
1 Gyr, the Earth will be sufficiently warm that a ``moist" greenhouse may occur and the temperature and composition profiles of the atmosphere will evolve so that the fraction of H$_{2}$O in the upper atmosphere is increased (Kasting 1988).   When the Sun's luminosity is 1.4 times its
current value, or when its age is 8 Gyr (see Girardi et al. 2000), then depending upon the effects of clouds, a runaway greenhouse may occur and the oceans will be totally vaporized (Kasting 1988).    As described by models for the early atmosphere of Venus, in the relatively water-rich upper atmosphere of the future, photodissociation of H$_{2}$O into OH and H (Ip 1983, Wu \& Chen 1993) will be a major source of ultraviolet opacity. Subsequently,  the resulting atomic hydrogen atoms may escape from the Earth in a wind.    Below, we extend this scenario for the Earth's future evolution to extra-solar terrestrial planets, and we argue that 
 in some circumstances, the wind of atomic hydrogen  from vaporizing oceans 
may produce 
observable absorption lines.  The Earth now is losing about 7 ${\times}$ 10$^{26}$ H atoms s$^{-1}$ (see, for example, Pierrard 2003);  we propose that this rate may increase by a factor of ${\sim}$10$^{3}$ in the future. 

\section{SCHEMATIC MODEL}

Because there are many uncertainties and unknowns, we adopt a simple, schematic model.  
  We assume oceans with 9.3 ${\times}$ 10$^{46}$ H nuclei on an analog to the Earth in a circular orbit at a distance, $D$, from the host star.    
 Once the planet is warm enough for a moist
greenhouse to occur,  much of the incident stellar ultraviolet  is absorbed by water in the upper atmosphere.  The photodissociated H$_{2}$O produces nonthermal hydrogen
atoms with a typical speed of 20 km s$^{-1}$ (Ip 1983, Wu \& Chen 1993) which
is greater than the escape velocity from the analog to the Earth of 11 km s$^{-1}$. However, because
the density in the environment where these photodissociations occur is sufficiently high, there are multiple collisions of the non-thermal hydrogen atoms with
ambient matter and a hydrodynamic treatment  of the outflow is appropriate. 

Here, we use a simple scaling of the detailed calculations by Watson et al.  (1981) and Kasting \& Pollack (1983). The fluid outflow rate from the planet is largely determined by the heating in the uppermost atmosphere which  is mainly composed of atomic hydrogen.  
  If $L_{EUV}$ is the star's luminosity in 
 hydrogen-ionizing photons, then the net outflow  of hydrogen atoms, ${\dot Z_{H}}$, is:
\begin{equation}
{\dot Z_{H}}\;{\approx}\;{\epsilon}_{wind}\frac{R_{pl}^{2}}{4\,D^{2}}\;\frac{L_{EUV}\,R_{pl}}{G\,M_{pl}\,m_{H}}   
\end{equation}
where ${\epsilon}_{wind}$ is the efficiency with which incident solar EUV photons are turned into heating the outflow, $R_{pl}$ and $M_{pl}$ and the radius and mass of the planet, and $m_{H}$ is the mass of a hydrogen atom.    For a region consisting largely of atomic hydrogen, we  adopt ${\epsilon}_{wind}$ = 0.3 (Shull 1979, 
Kasting \& Pollock 1983).    
Although  it varies by at least a factor of 3 during a solar cycle (see, for example, Ayres 1997),    a representative value for $L_{EUV}$ is ${\sim}$4 ${\times}$ 10$^{27}$ erg s$^{-1}$ (Judge, Solomon \& Ayres 2003, Ayres 1997),   thus yielding for an analog of the Earth that ${\dot Z}_{H}$ = 5 ${\times}$ 10$^{29}$ s$^{-1}$, coincidentally,  a typical molecular loss rate from a bright comet at 1 AU (Whipple \& Huebner 1976). 
Kasting \& Pollack (1983) have computed detailed models for the evolution of water in an atmosphere around Venus. When the lower atmosphere is largely composed of H$_{2}$O, their case D applies and the computed hydrogen mass loss rate is 1.2 ${\times}$ 10$^{30}$ s$^{-1}$.  Using its radius, mass and distance   from the Sun, we expect for Venus from equation (1) that ${\dot Z}_{H}$ = 1.1 ${\times}$ 10$^{30}$ s$^{-1}$,  consistent with much more sophisticated calculations.  We also note that with ${\dot Z}_{H}$ = 5 ${\times}$ 10$^{29}$ s$^{-1}$, the hydrogen wind persists for over ${\sim}$5 Gyr until the water-rich atmosphere produced by the vaporized oceans  is largely depleted.      

We now consider the detectability of this atomic hydrogen wind.  We assume that the wind flows outward at constant speed, $V$, and that  the atomic hydrogen is ionized with rate $J_{H}$ (s$^{-1}$).    
With the simplification of  spherical symmetry, the density, $n$, as a function of distance, $R$, from the
Earth is:
\begin{equation}
n(R)\;=\;\frac{{\dot Z_{H}}}{4{\pi}R^{2}V}\,exp\left(-\frac{R}{R_{0}}\right)
\end{equation}
where  $R_{0}$ is a characteristic length  (see, for example,  Jura \& Morris 1981)  such that
\begin{equation}
R_{0}\;=\;\frac{V}{J_{H}}
\end{equation}
Scaling from the solar system, we adopt  a photoionization rate of  1.6 ${\times}$ 10$^{-7}$ s$^{-1}$ at $D$ = 1 AU (Ayres 1997) and, including charge exchange with stellar wind protons (Combi \& Smythe 1988), a total 
hydrogen ionization rate at the Earth, $J_{0}$, of ${\sim}$10$^{-6}$ s$^{-1}$ or a mean
lifetime for neutral atoms of ${\sim}$10$^{6}$ s.  A characteristic bulk outflow
velocity of the hydrogen is 1 km s$^{-1}$ (Kasting \& Pollack 1983), but this value is uncertain and, as discussed below,  perhaps too low. The characteristic sizes
of a neutral hydrogen cloud around the terrestrial planet, $R_{0}$, are 1 ${\times}$ 10$^{11}$ cm ($V$ = 1 km s$^{-1}$) and $R_{0}$ = 5 ${\times}$ 10$^{11}$ cm ($V$ = 5 km s$^{-1}$).  In either case, the neutral hydrogen cloud  is  larger than the radius of the Sun  (7 ${\times}$ 10$^{10}$ cm), and therefore, during an occultation, there can be detectable absorption by neutral hydrogen atoms. 

Above, we assume that  the gas flows radially away from the planet at constant speed and is unaffected by any external forces.  In fact, radiation pressure by stellar Lyman ${\alpha}$
can accelerate the gas with rate, $a_{H}$ (cm s$^{-2}$).  If  $F_{\alpha}$ (erg cm$^{-2}$ s$^{-1}$ Hz$^{-1}$) denotes the flux in the center of the stellar 
line, then using cgs units, 
\begin{equation}
a_{H}\;=\;\left(\frac{F_{\alpha}}{m_{H}\,c}\right)\,\left(\frac{{\pi}e^{2}f}{m_{e}c}\right)
\end{equation}
where $e$ is the charge of the electron, $m_{e}$ the mass of an electron, $c$ the speed of light, and $f$ the oscillator strength of the absorption line (Spitzer 1978).
Extrapolating from the solar system, at 1 AU, $F_{\alpha}$ =  2.6 ${\times}$ 10$^{-12}$ erg cm$^{-2}$ s$^{-1}$ Hz$^{-1}$ (Combi \& Smythe 1988), and the resulting acceleration of a hydrogen atom  is 0.6 cm s$^{-2}$ since $f$ = 0.416  (Wiese, Glennon \& Smith 1966). If the atomic hydrogen cloud around the planet is  optically thin in the core of the Lyman ${\alpha}$ line,
then during the mean lifetime of a neutral hydrogen atom in the wind, it would accelerate to 6 km s$^{-1}$.  
Because the gas is optically thick, the actual dynamical effects of the photons are complex.   Here, because there are so many unknowns, we adopt the  drastic simplification that the outflow around the terrestrial planet  is  spherically symmetric, and that the effect
of radiation pressure from Lyman ${\alpha}$ is that   $V$ = 5 km s$^{-1}$.  In our discussion of the observability of the wind, we present results both for  
$V$ = 1 km s$^{-1}$ and $V$ = 5 km s$^{-1}$.       

With an estimate of the structure of the neutral hydrogen cloud, we can
now compute the strength and profile of the hydrogen absorption lines. 
If the cloud is
larger than the star, then at the moment when  the  occulting planet lies in front of the  center of the orbited star of radius $R_{*}$, the column density at  the limb of the star, $N(H)$, is:
\begin{equation} 
N(H)\;{\approx}\;\frac{{\dot Z_{H}}}{4\,R_{*}\,V}
\end{equation}
With $R_{*}$ = $R_{\odot}$, ${\dot Z_{H}}$ = 5 ${\times}$ 10$^{29}$ s$^{-1}$ and $V$ = 1 km s$^{-1}$ or 5 km s$^{-1}$,  then $N(H)$ 
= 1.8 ${\times}$ 10$^{13}$ cm$^{-2}$ or 3.6 ${\times}$ 10$^{12}$ cm$^{-2}$, respectively.   
Assuming a Gaussian line profile with a width characterized by ${\Delta}{\nu}_{0}$,  the optical depth at line center of an absorption line, ${\tau}(0)$, is 
\begin{equation}
{\tau}(0)\;=\;\frac{{\pi}e^{2}}{mc}\,f\,\;\frac{N(H)}{{\sqrt{\pi}}\,{\Delta}{\nu}_{0}}
\end{equation}
  For Lyman ${\alpha}$, we assume  Gaussian line profiles with ${\Delta}{\nu}_{0}$ = 8.23 ${\times}$ 10$^{9}$ Hz and ${\Delta}{\nu}_{0}$ = 4.11  ${\times}$ 10$^{10}$ Hz, corresponding to 1 km s$^{-1}$ and 5 km s$^{-1}$, respectively.    
With the column densities given above then ${\tau}(0)$ ${\approx}$ 14 ($V$ = 1 km s$^{-1}$) and ${\tau}(0)$ ${\approx}$ 0.5 ($V$ = 5 km s$^{-1}$).  As discussed below, such  
 lines are detectable if the velocity shift of the star
is sufficiently different from that of the  interstellar matter in the line-of-sight.  Higher order Lyman lines in the circumstellar gas might  be detectable even if observations at the stellar Lyman ${\alpha}$ velocity  are dominated   by intervening interstellar absorption.  We also note that
the absorption lines produced by the wind from the planet are likely to be narrower
than any structure in the the Lyman emission lines that are produced in the chromospheres of  main sequence G-type (Combi \& Smythe 1988, Redfield  et al. 2002) where the gas temperature is typically greater than 10,000 K and
even the thermal width of the hydrogen emission is greater than 10 km s$^{-1}$.   

We now consider how different levels of stellar activity can affect
the  observability of the  planet's wind.   
  The mass loss
rate of hydrogen from the planet (${\dot Z_{H}}$) scales with  $L_{EUV}$ and thus there is a tendency for more active stars to drive a stronger planetary wind and   produce
a larger projected column density of hydrogen.  However, an increase in the stellar activity also   leads to an increase in the neutral hydrogen ionization rate $(J_{H}$) and thus a  decrease in the size of the neutral cloud around
the planet ($R_{0}$).  To assess these combined affects, we compute the value of the
equivalent width of an absorption averaged over the disk of the star
when the planet lies toward the very center of the star.  We ignore any
limb brightening or darkening of the stellar atmosphere. Because there are so many uncertainties, we consider Doppler broadening parameters equal to the assumed outflow speeds of  either 1 km s$^{-1}$ or 5 km s$^{-1}$. 
  Considering main sequence stars of age, $t_{*}$, and  following Ayres' (1997) study of photo-ionization rates and Wood et al.'s (2002a) study of stellar winds, we assume  that both $J_{H}$ and $L_{EUV}$ scale as $t_{*}^{-2}$.   

We show in Figures 1 and 2 plots of $W_{\lambda}$ for the first three lines in the Lyman series vs. $J_{H}/J_{0}$. For both $V$ = 1 km s$^{-1}$ and $V$ = 5 km s$^{-1}$, we see that for $J_{H}/J_{0}$ ${\sim}$ 1, the predicted line strengths can be larger than 5 m{\AA} and therefore can be measured either
 with the Hubble Space Telescope or, possibly,  with the Far Ultraviolet Spectroscopic Explorer (see, for example, Meyer, Jura
\& Cardelli 1998, Wood et al. 2002b).     
We also see that for $J_{H}/J_{0}$ $>$ ${\sim}$20, which occurs when the star is younger than 1 Gyr,   the ionization rate of atomic hydrogen is so rapid  that the outflowing neutral cloud no longer effectively covers much of  image of the star
and the predicted equivalent widths decrease significantly.   This result suggests that
it will be difficult to use Lyman line absorptions to detect evaporated
oceans around stars younger than 1 Gyr.  Also, if $J_{H}/J_{0}$ $<<$ 1, the absorption lines are weak because not much atomic hydrogen is being lost from the planet.

\section{DISCUSSION}

We now consider observational strategies to find this
signature of evolved oceans.
  The probability that a star is transited by a planetary wind of characteristic size, $R_{0}$,   is
approximately $R_{0}/D$, which, for an analog to the Earth in orbit around a 
Sun-like main sequence star, is ${\sim}$7 ${\times}$ 10$^{-3}$ for V = 1 km s$^{-1}$ and ${\sim}$3 ${\times}$ 10$^{-2}$ for V = 5 km s$^{-1}$.  If the hydrogen-rich wind has a duration of 5
 Gyr, and the main sequence lifetime of the orbited star is about 10 Gyr, then there is a probability of 0.5 of observing the system during an era when  the planet has a detectable outflow.  
Since the local density of G-type main sequence stars is about 0.003 pc$^{-3}$ (see, for example, Greaves \& Wyatt 2003), then, depending upon the unknown fraction
of these stars that possess suitable  planets, a main sequence G-type star that  displays the transient hydrogen absorption
 may lie within ${\sim}$25 pc of the Sun.  Such stars are inside the local interstellar bubble and therefore the column of atomic hydrogen is typically less than 10$^{19}$ cm$^{-2}$ (Lehner et al. 2002).  In order to detect the transient hydrogen
lines during an occultation, the radial velocity of the star must be
sufficiently different from the interstellar velocity that the optical
depth in the damping wings of the interstellar hydrogen line is less than unity.  For a column
density of 10$^{19}$ cm$^{-2}$, the required velocity offsets  are
160 km s$^{-1}$, 33 km s$^{-1}$ and 12 km s$^{-1}$ for Lyman ${\alpha}$, Lyman
${\beta}$ and Lyman ${\gamma}$, respectively.  Since the typical velocity
dispersion of the G type stars in the solar neighborhood is 20 km s$^{-1}$, for most stars with transiting planets, it should be possible  to detect the associated transient circumstellar hydrogen absorption in
Lyman ${\gamma}$. Stars with lower column densities of interstellar matter and/or stars with somewhat higher
than average radial velocity may be suitable for observing at Lyman ${\beta}$ or, 
in the most favorable cases, Lyman ${\alpha}$.

When a terrestrial planet is found to occult a main sequence star, it should
be possible to infer from its period the distance of the planet from the star and therefore the planet's surface temperature and the amount of water that 
might have been evaporated.  Evolved oceans would be the most likely source of any sustained, substantial outflow of hydrogen from an extra-solar terrestrial planet.  

Our model of a hydrogen-rich wind can be tested.  (i)   The measured mass outflow rate in the hydrogen-rich wind, ${\dot Z}_{H}$, can be compared with the prediction from equation (1) after measuring the star's  X-ray emission.   (ii)  By observing the amount of hydrogen absorption before and after the star's occultation by the disk of the planet, it should be possible to test the characteristic spatial extent of the neutral gas, $R_{0}$, predicted by equation (2). To be exact, it will
be necessary to measure $V$ which can be done either directly from the line
profile or indirectly from the curve of growth.      
(iii) It may be possible to detect species besides hydrogen  in the outflow. For example,  
  in the Earth's oceans, [D]/[H] = 1.6 ${\times}$ 10$^{-4}$, about an order of magnitude greater than [D]/[H] within the interstellar medium.
In a planetary wind, some deuterium is  entrained with the outflowing hydrogen (see Keating \& Pollack 1983), and a [D]/[H] ratio significantly in excess of
the interstellar value could be another signature of an evolving ocean.  However, in our model, the predicted equivalent width of the deuterium Lyman ${\alpha}$ line
is $<$  0.1 m{\AA} and therefore very difficult to detect.

\section{CONCLUSIONS}
The oceans on a terrestrial planet may store the bulk of its hydrogen.  
When the host star's luminosity increases enough so that the oceans are substantially evaporated,  a 
 wind of atomic hydrogen from the planet could be strong enough to produce detectable transient  Lyman absorption lines in the star's ultraviolet spectrum.  Systems older than 1 Gyr 
are particularly promising candidates to exhibit  this signature 
of  terrestrial planets with evolved oceans.

The referee made insightful and helpful comments. 
This work has been partly supported by NASA.

\newpage
\begin{figure}
\epsscale{1}
\plotone{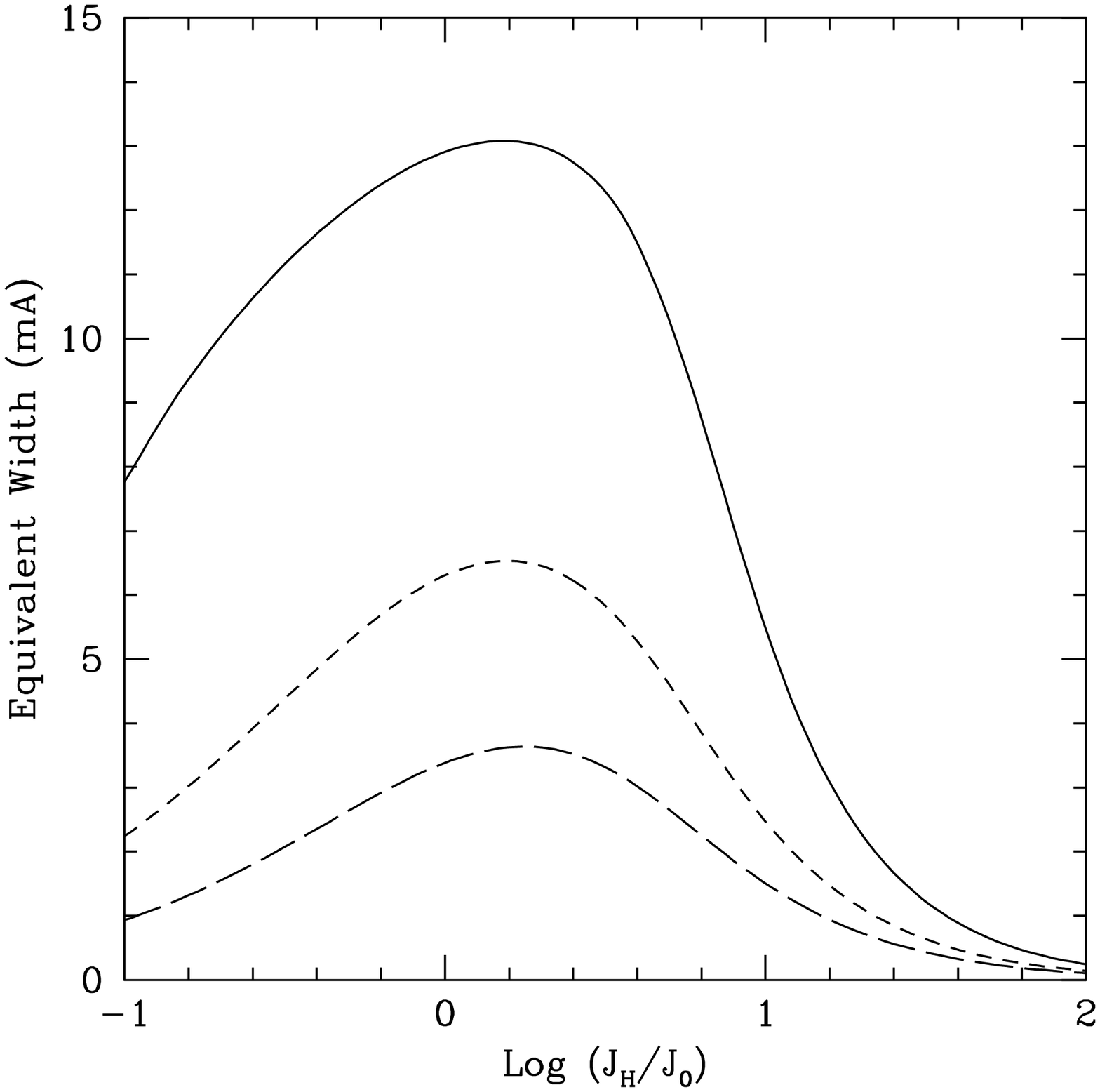}
\caption{Using the model described in the text, we show a plot of the predicted equivalent width of Lyman ${\alpha}$ (solid line), Lyman ${\beta}$ (short-dashed line) and Lyman ${\gamma}$ (long-dashed line) for different values of the relative hydrogen ionization rate, $J_{H}/J_{0}$.  As described in the text, we assume that $L_{EUV}$ scales as $J_{H}$. For these curves we assume an outflow velocity and a Doppler broadening parameter both equal to 1 km s$^{-1}$.}
\end{figure}
\begin{figure}
\epsscale{1}
\plotone{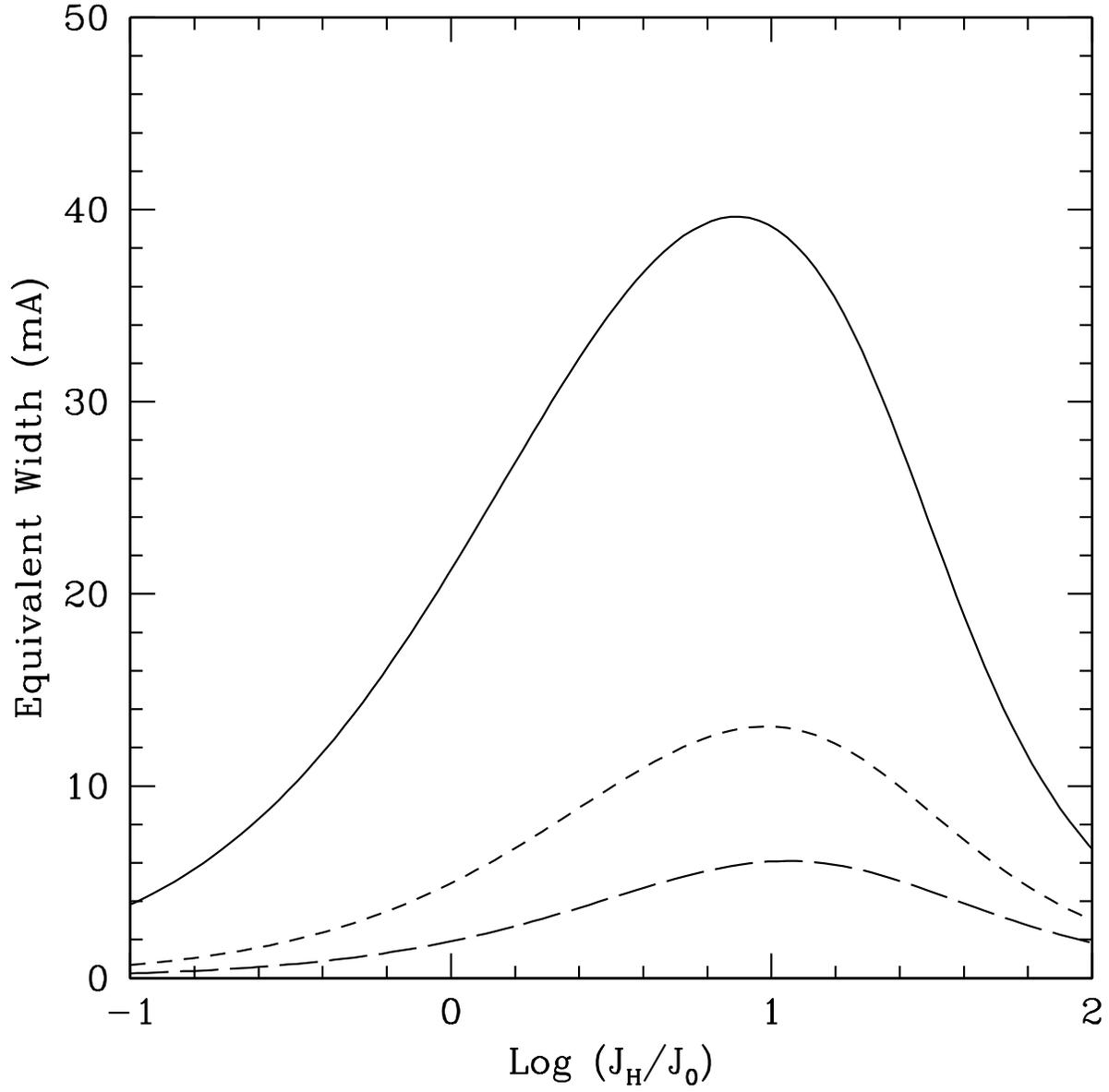}
\caption{The same as Figure 1 except that  we assume an outflow velocity and a Doppler broadening parameter both equal to  5 km s$^{-1}$.}
\end{figure}
\end{document}